\documentclass[12pt]{JHEP3}
\usepackage{epsfig}

\newcommand{\be}{\begin{equation}}
\newcommand{\ee}{\end{equation}}
\newcommand{\bea}{\begin{eqnarray}}
\newcommand{\beqa}{\begin{eqnarray}}
\newcommand{\eea}{\end{eqnarray}}
\newcommand{\eeqa}{\end{eqnarray}}
\newcommand{\ba}{\begin{array}}
\newcommand{\ea}{\end{array}}

\def\bbox{{\,
\lower0.9pt\vbox{\hrule \hbox{\vrule height 0.2 cm
\hskip 0.2 cm \vrule height 0.2 cm}\hrule}\,}}
\newcommand{\dsl}{\pa \kern-0.5em /}
\renewcommand{\th}{\theta}

\newcommand{\nn}{\nonumber \\}

\def\Tr{{\rm Tr\,}}

\def\ds{\raise.15ex\hbox{/}\kern-.57em\partial}
\def\Ds{\,\raise.15ex\hbox{/}\mkern-13.5mu D}
\def\Tr{{\rm Tr}\,}

\newcommand{\beq}{\begin{equation}}
\newcommand{\eeq}{\end{equation}} \newcommand{\beqn}{\begin{eqnarray}}
\newcommand{\eeqn}{\end{eqnarray}}

\newcommand{\CR}{\nonumber \\}
\newcommand{\pa}{\partial}

\newcommand{\D}{\delta}

\newcommand{\La}{\Lambda}
\newcommand{\p}{\Phi}

\newcommand{\da}{\dagger}

\preprint{
ITFA-2003-31\\
KIAS-P03047\\
hep-th/0306237\\
}

\title{\Large\bf Comments on Noncommutative Superspace}

\author{Seiji Terashima \& Jung-Tay Yee\footnote{Also at the Korea Institute
for Advanced Study, Seoul, Korea}\\
sterashi@science.uva.nl \& jungtay@science.uva.nl \\
\vskip 0.2mm
Institute for Theoretical Physics, University of Amsterdam \\
Valckenierstraat 65, 1018 XE Amsterdam, The Netherlands}

\abstract{ We study ${\cal N}={1 \over 2}$ supersymmetric theory
on noncommutative superspace which is a deformation of usual
superspace. We consider the deformed Wess-Zumino model as an
example and show the vanishing of vacuum energy, the
renormalization of superpotential and the non-vanishing of
tadpole. We find that the perturbative effective action has terms
which are not written in the star deformation. Also we consider
gauge theory on the noncommutative superspace and observe that
gauge group is restricted. We generalize the star deformation to
include noncommutativity between bosonic coordinates and fermionic
coordinates. }

\begin{document}

\section{Introduction}

Recently, field theories on noncommutative space have been
studied extensively.
These theories have some similarities to string theory, for example,
UV/IR mixing.
Indeed, noncommutative field theories can be derived from 
string theory by the Seiberg-Witten limit \cite{SeWi}. From
the dynamics of noncommutative field theories,
we expect to understand geometry behind string theory.

As a possible generalization of noncommutative field theories, 
we can supersymmetrize them. 
Supersymmetric field theories with noncommutativities only between
bosonic coordinates are easy to construct \cite{FeLl, Te}.
Then since, in supersymmetric theories, usual bosonic space is extended to
superspace which has fermionic coordinates, 
we are tempted to define noncommutative superspace as 
a further generalization and to construct 
field theories on the noncommutative superspace.
However, even though there have been many attempts \cite{Ca}-\cite{David},
the construction of 
the noncommutative superspace with non-anticommuting
fermionic coordinates
\beq
\{ \theta^\alpha, \theta^\beta \} = C^{\alpha \beta}
\eeq
with a constant symmetric deformation parameter  $C^{\alpha \beta}$
has been known to have serious difficulties.
To define analogues of (anti)chiral superfields,
we need to define supercharges $Q, \bar{Q}$
and covariant derivatives $D, \bar{D}$.
Those operators need to satisfy the Leibnitz rule,
which guarantees a product of chiral superfields is
again a chiral superfield.
However, it is extremely difficult to define such $Q,D$ in the
noncommutative superspace because some of them explicitly depend
on $\theta$.

Very recently, Seiberg showed that we can construct
Euclidean field theories on noncommutative superspace
and these field theories can be derived from the worldvolume 
theories of D-branes
in gravi-photon background \cite{Seiberg, BeSe}
\footnote{
It was first found
by Ooguri-Vafa \cite{Ooguri1,Ooguri2}
that string theories
in gravi-photon backgrounds
give rise to noncommutative superspaces.
}.
It is found out that these field theories
have only a half of supersymmetries compared with the field theories
without background or deformation. 

In this paper,
we consider classical and
perturbative quantum aspects of field theories defined 
on noncommutative superspace.
We find there are new kinds of (anti-)chiral superfields
in the ${\cal N}=\frac{1}{2}$ supersymmetric theories and
quantum fluctuations generate superpotential of those superfields 
in the effective action even though the original action
does not contain such terms.

The plan of this paper is as follows:
In section 2 we discuss
new kinds of (anti)chiral superfields in the
${\cal N}={1 \over 2}$ supersymmetric theory. We show that,  
although the notion of holomorphicity is violated,
the notion of anti-holomorphicity still survives 
in ${\cal N}={1 \over 2}$ supersymmetric theories.
This anti-holomorphicity leads to the
non-renormalization theorem of anti-superpotential
and the vanishing of vacuum energy for the deformed Wess-Zumino model.
We also consider gauge theories on the noncommutative superspace and
show that gauge group is restricted to products of $U(N)$.
We also show that $U(1)$ sector of $U(N)$ gauge group
is not decoupled from $SU(N)$ sector.
In section 3, we explicitly consider
perturbative dynamics of the deformed Wess-Zumino model.
We obtain terms like $\int d^2 \theta \Phi Q^2 \Phi$ with
divergent coefficient which was not
present in the original deformed Wess-Zumino model.
Section 4 is on discussions.

\section{ ${\cal N}=\frac{1}{2}$  supersymmetric theories}

According to \cite{Seiberg}\footnote{
We follow the notations of \cite{Seiberg}. In particular,
we use Lorentzian notations although we only consider
the Euclidean theory.},
we consider the following deformed superspace
\beqa &&
\{ {\hat{\theta}}^\alpha, {\hat{\theta}}^\beta \} = C^{\alpha \beta}, \;\;\;
\{ \bar{\theta}^{\dot{\beta}},{\hat{\theta}}^\alpha \}=
\{ \bar{\theta}^{\dot{\beta}}, \bar{\theta}^{\dot{\beta}} \}=
[ \bar{\theta}^{\dot{\beta}}, \hat{y}^\mu ] =0, \CR
&&
[  \hat{y}^\mu, \hat{y}^\nu ]=i \Theta^{\mu \nu}, \;\;\;\;
[  \hat{y}^\mu,{\hat{\theta}}^\alpha ]=\Psi^{\mu \alpha},
\label{deform}
\eeqa
where $\hat{\theta}$ and
$\hat{y}^\mu \equiv \hat{x}^\mu+
i \hat{ \theta}^\alpha  \sigma^\mu_{\alpha \dot{\alpha} }
\bar{\theta}^{\dot{\alpha}}$ are operators.
Note that $\theta$ is not complex conjugate of
$\bar{\theta}$.

A function of $\hat{\theta}, \hat{y}$ should be ordered.
In this paper we will always
use the Weyl ordered expression. Using a Fourier transformation,
it is written as
\beq
\hat{f}(\hat{y},\hat{\theta}) =
\int d^4 k \int d^2 \pi e^{-i k \hat{y} -\pi \hat{\theta} } \;
\tilde{f} (k,\pi).
\label{fhat}
\eeq
Then we can have a one to one map between a function of
$\hat{\theta}, \hat{y}$ to a function of ordinary
(anti)commutative coordinates $\theta, y$ via
\beq
f(y,\theta) =
\int d^4 k \int d^2 \pi e^{-i k y -\pi \theta } \;
\tilde{f} (k,\pi).
\label{fhat2}
\eeq

A product
$\hat{f_1}(\hat{y},\hat{\theta}) \, \hat{f_2}(\hat{y},\hat{\theta})$
is easy to compute:
\beq
\hat{f_1}(\hat{y},\hat{\theta}) \, \hat{f_2}(\hat{y},\hat{\theta}) =
\int d^4 k_1 d^4 k_2 \int d^2 \pi_1  d^2 \pi_2
e^{-i (k_1 +k_2) \hat{y} -(\pi_1 +\pi_2)  \hat{\theta} } \;
e^{i\Delta}
\tilde{f_1} (k_1,\pi_1) \tilde{f_2} (k_2,\pi_2),
\label{ff}
\eeq
where
\beq
e^{i\Delta}= e^{ \frac{1}{2} \left(
-\pi_1 C \pi_2
-i k_1 \Theta k_2
- k_1 \Psi \pi_2 + k_2 \Psi \pi_1 \right) },
\eeq
and
\beq
\pi_1 C \pi_2 =(\pi_1)_\alpha C^{\alpha \beta} (\pi_2)_\beta, \;\;\;
k_1 \Theta k_2 =(k_1)_\mu \Theta^{\mu \nu} (k_2)_\nu, \;\;\;
k_1 \Psi \pi_2=(k_1)_\mu \Psi^{\mu \alpha} (\pi_2)_\alpha.
\eeq
Now we define a star product between ordinary functions
in the momentum representation as follows :  
\beq
f_1(y,\theta) * f_2(y,\theta) \equiv
\int d^4 k_1 d^4 k_2 \int d^2 \pi_1  d^2 \pi_2
e^{-i (k_1 +k_2) y -(\pi_1 +\pi_2) \theta } \;
e^{i\Delta}
\tilde{f_1} (k_1,\pi_1) \tilde{f_2} (k_2,\pi_2).
\label{star}
\eeq
We can see that
$\hat{f_1}(\hat{y},\hat{\theta}) \, \hat{f_2}(\hat{y},\hat{\theta})$
is mapped to $f_1(y,\theta) * f_2(y,\theta)$.
By the change of integration variables $k=k_1+k_2, k'=k_1-k_2$ and
$\pi=\pi_1+\pi_2, \pi'=\pi_1-\pi_2$,
(\ref{ff}) becomes a form of (\ref{fhat}) from which
we can identify a corresponding function $f_1 *f_2$ of ordinary coordinates.

We can see the star product is
associative just from the definition as in the usual Moyal star product
for the bosonic noncommutativity ($C=\Psi=0$ in our notation).
For $\Theta=\Psi=0$ case,
the star product is the same as the
fermionic star product defined in \cite{AsSuTe} \cite{Seiberg}, which is given
by
\beq
f_1(\theta) * f_2(\theta) =
\left.
e^{-\frac{1}{2} C^{\alpha \beta}
\frac{\partial}{\partial \theta_1^\alpha}
\frac{\partial}{\partial \theta_2^\beta}
}
f_1(\theta_1) f_2(\theta_2)
\right|_{\theta_1=\theta_2=\theta}.
\label{star1}
\eeq
In general, the star product becomes
\beq
f_1(y,\theta) * f_2(y,\theta) =
\left.
 e^{ \frac{1}{2} \left(
-\frac{\partial}{\partial \theta_1}  C
\frac{\partial}{\partial \theta_2}
+i \frac{\partial}{\partial y_1} \Theta
\frac{\partial}{\partial y_2}
- i  \frac{\partial}{\partial y_1} \Psi
\frac{\partial}{\partial \theta_2}  +
i \frac{\partial}{\partial y_2}  \Psi
\frac{\partial}{\partial \theta_1} \right) }
f_1(y,\theta_1) f_2(y, \theta_2)
\right|_{y_1=y_2=y, \; \theta_1=\theta_2=\theta}.
\label{star2}
\eeq

Now we can construct field theories on the
noncommutative superspace by just replacing ordinary products of
superfields to the star products (\ref{star2}).
But we should be a little bit more careful about 
supercharges $Q, \bar{Q}$ and covariant derivatives $D, \bar{D}$.
In $y, \theta, \bar{\theta}$ coordinates system, we are allowed to define
supercharges and covariant derivatives as 
\beqa
Q_\alpha &=&
 { \partial \over \partial
 \theta^{\alpha}}, \CR
\bar{Q}_{\dot{\alpha}} &=&
-{\partial \over \partial \bar{\theta}^{\dot{\alpha}}}
 +2i \theta^{\alpha}
\sigma^\mu_{\alpha \dot{\alpha}}
 {\partial \over \partial y^\mu}, \CR
D_\alpha &=& {\partial \over \partial \theta^\alpha}
 +2i\sigma^\mu_{\alpha \dot{\alpha}} \bar{\theta}^{\dot{\alpha}}
 {\partial \over \partial y^\mu}, \CR
\bar{D}_{\dot{\alpha}}
&=& - { \partial \over \partial
 \bar{\theta}^{\dot{\alpha}}}.
\eeqa
However, we should note that  
$\bar{Q}_{\dot{\alpha}}$
is not a linear operator and it does not generate 
a symmetry. The reason is because $\theta$ in the definition can not
be considered as a constant any more. 

We also can define chiral and anti-chiral superfields in the 
star product formalism. 
Chiral superfields are defined as
$\bar{D}_{\dot{\alpha}} \Phi(y,\theta,\bar{\theta})=0$,
which can be rewritten as $\Phi(y,\theta,\bar{\theta})=\Phi(y,\theta)
=A(y)+\sqrt{2} \theta \psi(y) +\theta \theta F$.
Anti-chiral superfields are defined as
$D_{\alpha} \bar{\Phi}(y,\theta,\bar{\theta})=0$,
which again are written in terms of component fields as
$\bar{\Phi}(y,\theta,\bar{\theta})=
\bar{\Phi}(\bar{y},\bar{\theta})
=\bar{A}(\bar{y})+\sqrt{2} \bar{\theta} \psi(\bar{y})
+\bar{\theta} \bar{\theta} \bar{F}$.
Here $\bar{y} \equiv y^\mu -2 i \theta^\alpha
\sigma^\mu_{\alpha \dot{\alpha}}  \bar{\theta}^{\dot{\alpha}}$.
There is no ambiguity in the above expressions
of $\Phi(y,\theta)$ and $\bar{\Phi}(\bar{y},\bar{\theta})$. From
these superfields, we can always construct a
Lagrangian as
\beq
L= \int d^4 \theta K(\Phi,\bar{\Phi})_*
+\int d^2 \theta  W(\Phi)_* |_{\bar{\theta}=0}
+\int d^2 \bar{\theta} \bar{W}(\bar{\Phi})_*|_{\theta=0}.
\label{lag}
\eeq
All products are star products. The action is given by $\int d^4 x L(x)$.
Here we note that
$\int d^4 x Q(x,\theta,\bar{\theta})= \int d^4 x Q(y,\theta,\bar{\theta})$
since the difference
is the integration of a total divergence.

\subsection{Chiral superfields in ${\cal N}=1/2$ superspace}

In this subsection, we first forget about the star product and
consider ${\cal N}=1/2$ supersymmetric theory
on usual (anti)commutative superspace.
From a chiral superfield $\Phi$, we
can construct other chiral superfields by multiplication,
$\Phi * \Phi$ or by differentiation $\partial_\mu \Phi$.
Interestingly, if we keep only the ${\cal N}=1/2$ supersymmetry,
we can construct new kinds of chiral superfield from $\Phi$.
$Q_\alpha  \Phi$ and $Q^2 \Phi$
are chiral super fields since
$\bar{D}_{\dot{\beta}} ( Q_\alpha \Phi)=0$
and $Q_\beta (Q_\alpha \Phi)= - Q_\alpha (Q_\beta \Phi)$.
From the anti-chiral superfield $\bar{\Phi}$,
we can construct anti-chiral superfields
$Q \bar{\Phi}, \bar{\theta}_{\dot{\beta}}  \bar{\Phi}$ and
$\bar{\theta} \bar{\theta}  \bar{\Phi}$.
They vanish by acting $D$ on them
and $\{ Q, \bar{\theta} \}=\{ Q, Q \}=0$.
Other chiral or anti-chiral superfields can be rewritten in terms 
of these basic superfields. For example, $Q^2 \bar{\Phi}$ 
can be rewritten using $\bar{\theta} \bar{\theta} \bar{\Phi}$
as $Q^2 \bar{\Phi} \sim \Box (\bar{\theta} \bar{\theta} \bar{\Phi})$.
There are no other new kinds of chiral nor anti-chiral superfields.

Now we briefly mention several novel features of the
${\cal N}=1/2$ superspace.
An interesting property of ${\cal N}=1/2$ superfields is that
$F$ and $\bar{A}$ are invariant themselves under
the ${\cal N}=1/2$ supersymmetry transformation, 
which is generated by $Q$.
This property is related to the fact that
$F=Q^2 \Phi$ and
$\bar{\theta} \bar{\theta}\bar{A}=\bar{\theta} \bar{\theta} \bar{\Phi}$
are superfields.
Another interesting feature of the ${\cal N}=1/2$ superfields is that
there are supersymmetric invariants without the integration of spacetime.
Polynomials constructed from arbitrary multiplications of
$ \left(\int d^2 \theta G(\Phi_i) \right)|_{\bar \theta=0}$ and
$\left( \int d^2 \bar\theta (\bar\theta \bar\theta )
\bar G (\bar\Phi_j) \right)$ are supersymmetric invariants
since the supertransformation of these does not have derivative terms.
Then using these polynomials,
we can construct ${\cal N}=1/2$ supersymmetric action
which contains terms with arbitrary number of $\int d^2 \theta$
and $\int d^2 \bar{\theta}$. Or if this type of construction 
looks unsatisfactory, since there are almost trivial identities
\beqa
&&\left(\int d^2 \theta G(\Phi_i(y,\theta)) \right)|_{\bar \theta=0}
\left(\int d^2 \theta' G'(\Phi_i(y,\theta')) \right)|_{\bar \theta=0} \nn
&&=\left(\int d^2 \theta
(Q^2 G(\Phi_i(y,\theta))) G'(\Phi_i(y,\theta)) \right)|_{\bar \theta=0}, \nn
&& \left(\int d^2 \bar \theta (\bar\theta \bar\theta )
\bar G(\bar \Phi_i(\bar y,\bar \theta)) \right)
\left(\int d^2 \bar \theta' (\bar\theta' \bar\theta' )
\bar G'(\bar \Phi_i(\bar y,\bar \theta')) \right) \nn
&&=
\int d^2 \bar \theta (\bar\theta \bar\theta )
\bar G(\bar \Phi_i(\bar y,\bar \theta))
\bar G'(\bar \Phi_i(\bar y,\bar \theta)) , \nn
&&
\left(\int d^2 \bar \theta (\bar\theta \bar\theta )
\bar G(\bar \Phi_i(\bar y,\bar \theta)) \right)
\left(\int d^2 \theta' G'(\Phi_i(y,\theta')) \right)|_{\bar \theta=0} \nn
&&=\int d^2 \theta d^2 \bar \theta (\bar\theta \bar\theta )
\bar G(\bar \Phi_i(\bar y,\bar \theta))
G'(\Phi_i(y, \theta)), \nonumber
\eeqa
we are allowed to  
change a product of these unusual supersymmetric invariants
to a single integration of $\theta$ and $\bar{\theta}$. Anyway, 
note that this kind of construction is not possible
for usual ${\cal N}=1$ supersymmetric theories, but unique to ${\cal N}=1/2$
theories.

In summary, we can construct ${\cal N}=1/2$ supersymmetric Lagrangian on 
commutative superspace as follows.
First, we consider a Lagrangian on usual
${\cal N}=1$ commutative superspace and next
add terms (for example, $\mu \int d^2 \bar{\theta} \;
\bar{\theta} \bar{\theta} \bar{\Phi}^2=\mu \bar{A}^2$) constructed with 
new kinds of chiral and anti-chiral superfields 
to the ${\cal N}=1$ Lagrangian.
Since additional terms break ${\cal N}=1$ supersymmetry
to ${\cal N}=1/2$ supersymmetry,
we have ${\cal N}=1/2$ supersymmetric theory on
usual (anti)commutative superspace.
Generically, this ${\cal N}=1/2$ supersymmetric theory
has no relation to the deformation (\ref{deform}) nor
the noncommutative superspace.

Now let us consider the Lagrangian on the noncommutative superspace
(\ref{lag}).
It is important to notice that the star products (\ref{star1}) and
(\ref{star2}) can be written in terms of
$Q_\alpha (=\frac{\partial}{\partial \theta^\alpha} )$
and $\partial_\mu$.
This means that, using those new kinds of superfields,
we can rewrite (\ref{lag}) to
the Lagrangian on usual (anti)commutative superspace
without the star product.
For example, the $C^{\alpha \beta}$ dependent term of
the Wess-Zumino Lagrangian in \cite{Seiberg}
is given by
\beq
\frac{1}{3} g \, {\rm det} C F^3=
\frac{1}{48} g \, {\rm det} C
\int d^2 \theta \Phi (Q^2 \Phi) (Q^2 \Phi).
\label{cl}
\eeq
Thus the set of
the theories on noncommutative superspace
is considered as
a special subset of ${\cal N}=1/2$ supersymmetric theory.

We will see later that quantum fluctuations of
the deformed Wess-Zumino model of \cite{Seiberg} generate 
the term  $\int d^2 \theta \Phi Q^2 \Phi$ to the
effective Lagrangian.
This term can not be obtained from the star deformation of
any ${\cal N}=1$ supersymmetric theory.
The question about the characterization of 
noncommutative superspace in terms of ${\cal N}=1/2$ superspace
will be left as a future problem.

Another interesting question we can address is the non-renormalization 
theorem for ${\cal N}=1/2 $ supersymmetric theories.
If we report the result in advance, 
non-renormalization theorem and anti-holomorphicity
survives for anti-superpotential, but holomorphicity 
is violated for superpotential and superpotentials are renormalized.
To prove the non-renormalization theorem
for ${\cal N}=1 $ supersymmetric theories,
we can use the argument by Seiberg and Intriligator\cite{SeIn} : 
promote coupling constants to
chiral or anti-chiral superfields and use the notion of holomorphicity.
In order to apply this argument to ${\cal N}=1/2$ supersymmetric theories,
it is important to note that
the superpotential $W$ is not distinguishable from
the K\"{a}hler potential $K$.
This is because $ \bar{\theta}\bar{\theta} \sim
\delta (\bar{\theta})$ is an anti-chiral superfield
and a K\"{a}hler term $\int d^4 x \int d^4 \theta \,\,
(\bar{\theta}\bar{\theta})
\tilde{K}(\Phi(y,\theta),\bar{\Phi}(\bar{y},\bar{\theta})) $
can be converted to a superpotential
 $\int d^4 x \int d^2 \theta
\tilde{K}(\Phi(x,\theta),\bar{A}(x))$.
This feature is related to the fact that
$\bar{A}(x)$ is invariant under the supersymmetry transformation.
The lowest component of the anti-chiral superfield $\bar{A}$ itself
can appear in the superpotential. Consequently 
we expect the non-renormalization theorem
for the superpotential is no longer valid.
Indeed we will see explicitly in the next section that
there are quantum corrections to
the superpotential of the deformed Wess-Zumino model.
On the other hand, the anti-superpotential is 
distinguished from the  K\"{a}hler potential  and the superpotential.
Thus we can safely use the notion of anti-holomorphicity
for anti-superpotential.

Another interesting quantity  to consider in supersymmetric theories 
is vacuum energy.
Since we have no $\bar{Q}$
in the ${\cal N}=1/2$ supersymmetric theory,
we can not conclude vacuum energy is zero
from the algebraic relation $P=\{ Q, \bar{Q} \}$.
However, we can argue the deformed Wess-Zumino model
has vanishing vacuum energy by the still-remaining
  ${\cal N}=1/2$ supersymmetry.
This is because vacuum energy is represented by
an anti-superpotential
\beq
\int d^4 y \int d^2 \bar{\theta} (\bar{\theta}\bar{\theta})
\Lambda_0 = \int d^4 y \Lambda_0.
\eeq
The deformation parameter $C^{\alpha \beta}$
enters into the Lagrangian as $g {\rm det }C$.
Hence
$g$ and $ C^{\alpha \beta}$
can be considered as the lowest components of chiral superfields
which can not appear in the anti-superpotential.
Considering the  the fact
that the vacuum energy is zero for $C^{\alpha \beta}=0$ case, 
we can argue that the vacuum energy of the deformed Wess-Zumino model must
be zero in all order in perturbation theory.
Furthermore, we can conclude that
anti-superpotential of the deformed Wess-Zumino model
is not renormalized since $C^{\alpha \beta}$
can be considered as the lowest component of a chiral superfield.

\subsection{Gauge theory on noncommutative superspace and
restriction of the gauge group}

Now we consider vector superfields in the noncommutative
superspace following \cite{Seiberg}.
The gauge symmetry acts on the vector superfields as
\beq
(e^V)_* \rightarrow (e^{V'})_* = (e^{-i \bar{\Lambda}})_*
 * (e^V)_*  * (e^{i \Lambda})_*,
\eeq
where $*$ in the $(e^V)_*$ means that all the product in the exponential
are understood as the star products.
The chiral and anti-chiral field strength superfields
are given by
\beqa
W_\alpha &=& -\frac{1}{4} \bar{D}\bar{D} (e^{-V})_*
* D_\alpha (e^{V})_* \CR
\bar{W}_{\dot{\alpha}} &=& \frac{1}{4} D D (e^{V})_*
* \bar{D}_{\dot{\alpha}} (e^{-V})_*.
\eeqa
The gauge transformation for them are
\beqa
W_\alpha & \rightarrow &
(e^{-i \Lambda})_* * W_\alpha * (e^{i \Lambda})_*,
\CR
\bar{W}_{\dot{\alpha}} &\rightarrow&
(e^{-i \bar{\Lambda}})_* * \bar{W}_{\dot{\alpha}}
* (e^{i \bar{\Lambda}})_*.
\eeqa
From these superfields, we can construct the
Lagrangian on noncommutative superspace \cite{Seiberg}.

We regard $V$ as a matrix valued vector superfield 
in order to consider non-Abelian gauge theory.
In this case, the gauge group is restricted by the requirement 
of the consistency of the gauge transformation
as is the case in the bosonic noncommutative gauge theory \cite{Te}.
Let us be more specific. 
We denote $W_\alpha=T^a W_\alpha^a$, where $T^a$ is a $ d_r \times d_r$ matrix
for a representation $R$ of some gauge group $G$ and satisfies
$(T^a)^\da=T^a$ and $\Tr (T^a T^b)=k$.
The infinitesimal version of the gauge transformation is
\beq
\D W_\alpha \!\!
= - i [\Lambda, W_\alpha]
= -\frac{i}{2} [ T^a, T^b ] (\Lambda^a \!*\! W_\alpha^b
+ W_\alpha^b\!*\!\Lambda^a  )
-\frac{i}{2} \{ T^a, T^b \}
(\Lambda^a \!*\! W_\alpha^b - W_\alpha^b\!*\! \Lambda^a ).
\label{igt}
\eeq
Note that the last term does not vanish because of non-(anti)commutativity.
Therefore the gauge transformation does not close and it is inconsistent
if  $\{ T^a, T^b \}$ is not a linear combination of
$T^d$ for some $a,b$. This means $T^a$ spans a basis
of the complex algebra of the $ d_r \times d_r$ complex matrices.
Thus it is obvious that the gauge transformation is consistent
only for the unitary group $G=U(N)$ or its direct product
$G=\prod_a U(N_a)^{(a)}$.
Here
we should take $T^a$ as the matrix for
the fundamental or anti-fundamental representation.
Indeed the fundamental or anti-fundamental representation of the $U(N)$
spans the whole space of $N \times N$ complex matrices.

Moreover,
the representations of the gauge group $G$ of the matter is also
restricted to fundamental ($\bf N$), anti-fundamental ($\bar{\bf N}$),
adjoint (${\bf N} \times {\bar{\bf N}}$)
or bi-fundamental (${\bf N} \times {\bar{\bf M}}$)
from the consideration of the possible form of
the K\"{a}hler potential of the matter.
This is because even though we can use 
$T^a$ of the fundamental representation for the matter,
just as in the bosonic noncommutativity \cite{Te},
we can not construct a gauge invariant K\"{a}hler potential
for the other tensor representations, for example $({\bf N} \times {\bf N})$.
The gauge transformations for the fundamental, anti-fundamental,
adjoint and bi-fundamental chiral superfields are given by
\beqa
\p & \rightarrow & (e^{-i \La})_* * \p, \CR
\tilde{\p} & \rightarrow &  \tilde{\p} * (e^{i \La})_*, \CR
\p_{adj} & \rightarrow &
(e^{-i \La})_* * \p_{adj} * (e^{i \La})_* , \CR
\p_{N\bar{M}} & \rightarrow &
(e^{-i \La^{(1)}})_* * \p_{N\bar{M}} * (e^{i \La^{(2)}})_* ,
\eeqa
respectively
and the Lagrangians are given by
\beqa
L_{\p} &=&   \int d^4 \th
\left( \bar{\p} * (e^{-V})_* * \p \right) \CR
L_{\tilde{\p}} &=&   \int d^4 \th
\left( \tilde{\p} * (e^{V})_* * \bar{\tilde{\p}} \right) \CR
L_{\p_{adj}} &=&  \int d^4 \th
\frac{1}{k} \Tr
\left( (e^{V})_* * \bar{\p}_{adj}
* (e^{-V})_* * \p_{adj} \right) \CR
L_{\p_{N\bar{M}}} &=& \int  d^4 \th
\Tr \left(
(e^{V^{(2)}})_* * \bar{\p}_{N\bar{M}} * (e^{-V^{(1)}})_*
* \p_{N\bar{M}} \right).
\label{S4}
\eeqa

For
$\{ {\hat{\theta}}^\alpha, {\hat{\theta}}^\beta \} =
C^{\alpha \beta}, \;
[  \hat{y}^\mu, \hat{y}^\nu ]=
[  \hat{y}^\mu,{\hat{\theta}}^\alpha ]=0$ case,
we can also see this gauge group restriction from 
the $C^{\alpha \beta}$-dependent terms in the gauge fixed
Lagrangian of the component fields \cite{Seiberg}
\beq
-i C^{\mu \nu} \, \Tr F_{\mu \nu} \bar{\lambda}\bar{\lambda}
+ {\rm det} C \, \, \Tr (\bar{\lambda}\bar{\lambda})^2
\label{lagg}.
\eeq
Here $F_{\mu \nu}$ and $\bar{\lambda}$ are transformed
as adjoint representations of the gauge group.
However, since the couping of (\ref{lagg}) is not written
by commutators, 
to cancel the contribution from the gauge transformation of (\ref{lagg}),
the gauge transformation of $\lambda$  should have 
a term proportional to $\bar{\lambda}\bar{\lambda}$ \cite{Seiberg}.
It is inconsistent for any gauge group except
$G=\prod_a U(N_a)^{(a)}$.

Related to this gauge group restriction,
we can see that the $U(1)$ sector of the $U(N)$ group
is not decoupled from the $SU(N)$ sector since
the second term of (\ref{lagg}) contains
$\Tr (( \bar{\lambda}_{U(1) }\bar{\lambda}_{U(1)})
(\bar{\lambda}_{SU(N) }\bar{\lambda}_{SU(N)}))=
\bar{\lambda}_{U(1) }\bar{\lambda}_{U(1)}
\Tr (\bar{\lambda}_{SU(N) }\bar{\lambda}_{SU(N)}) $
which is not zero.
Here $\bar{\lambda}_{U(1)}$ and $\bar{\lambda}_{SU(N) }$ are
the $U(1)$ part and the $SU(N)$ part of the $\bar{\lambda}$ respectively.
Furthermore, we find out the $U(1)$ gauge theory without matter
is not trivial.
This is because the coupling $\Tr F_{\mu \nu} \bar{\lambda}\bar{\lambda}
=F_{\mu \nu} \bar{\lambda}\bar{\lambda}$
does not vanish even for $U(1)$ group, even though 
all fields are adjoint representations of the $U(1)$,
i.e. chargeless.

As we did on the deformed Wess-Zumino model, we are tempted to
reformulate gauge theories on noncommutative superspace as
theories on (anti)commutative superspace, 
since the star deformations contains $Q$ and $\partial$ only
even if we include the vector multiplets into the Lagrangian.
However, the gauge symmetry on the noncommutative superspace
is different from the gauge symmetry on the (anti)commutative superspace.
In the component fields formulation, the two gauge transformations have the
same form only
after the redefinition of the components fields of the noncommutative
vector superfield \cite{Seiberg}.
This means two gauge symmetry are indeed different.
Therefore, the gauge symmetry is not manifest
if we simply rewrite gauge theories on noncommutative superspace
as gauge theories on (anti)commutative superspace.

Now we briefly comment about Lorentz symmetry, which  is
the symmetry of rotation of $x^i$ $(i=1,\cdots,4)$.
It is $SO(4)=SU(2)_L \otimes SU(2)_R$ in usual commutative Euclidean space.
But, in the noncommutative superspace,
one $SU(2)_L$ sector of the Lorentz symmetry corresponding to
undotted $SU(2)_L$ indices is broken because of the presence of
$C^{\alpha \beta}$.  Note that another $SU(2)_R$ sector remains unbroken.
We can say that since the supercharges commute with
both the translation operators $P_\mu$
and the unbroken $SU(2)_R$ generator of the Lorentz symmetry,
the ${\cal N} =1/2$ supersymmetry is decoupled from
the space-time symmetry for a generic value of $C^{\alpha \beta}$.
However, if we set $C^{\alpha \beta}$ to some special value, 
$U(1)$ subgroup of the $SU(2)_L$
remains unbroken and the supersymmetry and the space-time symmetry
are not decoupled.
For example, if we set $C^{\alpha \beta}=\delta_{\alpha \beta}$,
$U(1)$ subgroup $O(a)=\exp( i a \sigma_2) \in SU(2)_L$ 
where $a$ is a real parameter is unbroken. Since
$O(a)$ is real as we see from $O(a)^T=O(a)^{-1}$, 
$C^{\alpha \beta} (=\delta_{\alpha \beta})$ is invariant under
the transformation generated by
$O(a)$ as $O(a) \delta_{\alpha \beta} O(a)^T =\delta_{\alpha \beta} $.
Actually, this is nothing but a rotation
between $\theta^1$ and $\theta^2$.

Finally we consider the matrix model which is formally equivalent to
a field theory on the noncommutative superspace
with $C^{\alpha \beta}$ and
a non-degenerate noncommutative parameter
$\Theta^{\mu \nu}$.
The action of the matrix model
is given from any field theory on the noncommutative superspace
by the following way.
First, using one to one map of (\ref{fhat}) and (\ref{fhat2}), 
we replace any superfield $G(y,\theta,\bar{\theta})$ in the action
to a matrix which depends on anti-commuting parameter $\bar{\theta}$,
$\hat{G} (\hat{y},\hat{\theta},\bar{\theta} )=
\hat{G}_1 (\hat{y},\hat{\theta})+ \hat{G}_2 (\hat{y},\hat{\theta}) \bar{\theta}
+\hat{G}_3 (\hat{y},\hat{\theta}) \bar{\theta}\bar{\theta}$.
Here we regard the operator $\hat{y}$ as an infinite dimensional matrix
as we do in the usual bosonic noncommutative field theory and
$\hat{\theta}^\alpha$ as $\gamma^\alpha (=\sigma_\alpha)$ 
which are two dimensional gamma matrices.
Now the $\hat{G}_i(\hat{y},\hat{\theta})$ is a $(2 \infty) \times (2 \infty)$
matrix.
The integration $\int d^4 y$ is replaced by a trace
for $\hat{y}$, $\Tr_{\hat{y}}$.
As shown in \cite{AsSuTe}, we can replace $\int d^2 \theta$ and
$\left( \cdots \right) |_{\theta=0} $
to the supertrace of the gamma matrices,
$ \frac{i}{4}  \Tr_{\gamma} \left(\sigma_3 \cdots \right)$, and
the trace of the gamma matrices, $\frac{1}{2} \Tr_{\gamma}$, respectively.
Then finally we obtain a (super)matrix model.
Note that we should consider $\gamma_\alpha$ as a fermion, i.e.
anti-commutes with $\bar{\theta}$ or fermionic component fields.

\section{Quantum Aspects of the Wess-Zumino Model in 
Noncommutative Superspace}

In this section to see the quantum properties discussed in
the previous section clearly,
we study the Wess-Zumino model in the noncommutative superspace
\bea
\label{original}
 L = \int d^4 \theta \Phi \bar\Phi + \int d^2 \theta
  \left({m\over 2} \Phi * \Phi + {g \over 3} \Phi*\Phi*\Phi \right)
 + \int d^2 \bar\theta
  \left( {\bar m\over 2} \bar\Phi * \bar\Phi
  + {\bar g\over 3} \bar\Phi* \bar\Phi * \bar\Phi \right). \nn
\eea
We set $\Theta=\Psi=0$ to separate the effect of 
fermionic noncommutativities only.
In component fields formulation, the effect of the star deformation
is to add $F^3$ term \cite{Seiberg}, which means only the scalar potential
is affected by the deformation.
The potential expressed in components fields is
\be
\label{potential}
 V = -F\bar F - m A F -g A^2 F-  {g \over 3} \det C F^3
    - \bar m \bar A \bar F - \bar g \bar A^2 \bar F.
\ee
To eliminate the auxiliary fields $F$ and $\bar F$, we need equations of
motion
\bea
 &&\bar F + m A + g \det C F^2 + g A^2 =0,  \nn
 &&F+ \bar m \bar A + \bar g \bar A^2 =0.
\eea
Then the potential (\ref{potential}) is expressible as
\be
\label{potentialwithoutF}
 V = V(C_{\alpha\beta}=0) +
     \det C\left({1\over 3} g \bar m^3 \bar A^3 + g\bar g \bar m^2
    \bar A^4 + g\bar g \bar m \bar A^5 + {1 \over 3}
     g\bar g^3 \bar A^6 \right).
\ee

In this paper, since $\bar g$ does not appear in the N-point
functions of $\Phi$s at 1-loop level,
we take  $\bar g \rightarrow 0$ limit
to understand the essential physics of noncommutative
superspace avoiding unnecessary complexities.
Then since $\bar\Phi$ is free,
 we can integrate out $\bar \Phi$ from (\ref{original}),
leaving only terms containing $\Phi$ \cite{Zanon}.
\bea
\label{action}
 S = \int d ^4 y d ^2 \theta \left( {1 \over 2} \Phi(y,\theta)
  ( m - {\square
  \over \bar m})  \Phi(y,\theta) +
    {1 \over 3} g \Phi(y,\theta) * \Phi(y,\theta) * \Phi(y,\theta) \right).
\eea
Note that $\int \Phi * \Phi = \int \Phi \Phi$.
Since the representation of $*$-operation in  momentum superspace
is much simpler than its  position space representation,
it is convenient to use momentum space Feynman rules to calculate
quantum corrections. Superfield in momentum space is defined as
\be
 \Phi(p, \pi) = \int d^4y d^2 \theta \exp(ipy+\pi\theta)  \Phi(y, \theta).
\ee
Expanded into components fields, the momentum space
superfield is expressible as
\be
 \Phi(p, \pi) = {1 \over 4} \pi \pi A(p) + {1 \over \sqrt{2}} \pi \psi(p)
  + F(p).
\ee

The quantization of the action (\ref{action}) is straightforward.
 One nice property of the action (\ref{action}) is that separate
treatment between bosonic coordinates and fermionic coordinates is
possible. Since we have only $\Phi(y,\theta)$ in the action,
we don't need chiral projectors any more.
For bosonic coordinates, Feynman rules are nearly the same as the scalar
$\phi^3$ theory. One slight modification needed is that we should use
\be
 {\bar m \over p^2 + m \bar m}
\ee
as a propagator.
For fermionic coordinates,
using the standard Wick-contraction procedure, we get two kinds of
interaction vertices in the momentum superspace : Twisted
and untwisted vertices.
Via Fourier transformation, the Feynman rule is simply to attach a phase factor
\be
 \exp(-{C^{\alpha\beta} \over 2} \pi_{1\alpha} \pi_{2\beta})
 \equiv \exp(-{1 \over 2}\pi_1 \wedge \pi_2),
\ee
for an untwisted vertex and
\be
 \exp(+{1 \over 2}\pi_1 \wedge \pi_2),
\ee
for a twisted vertex. But the classification of twisted
and untwisted vertex
is just a relative matter as in the case of bosonic noncommutativity.
As we will see from below, interesting new physics
arises from the sectors containing nonplanar diagrams.
This is because the planar diagrams do not depend
on the deformation parameter except for
the star products between the external legs \cite{Minwalla}.

\subsection{Vacuum energy}

One loop vacuum diagram vanishes since the diagram is planar.

\begin{figure}[h]
\begin{center}\mbox{\epsfysize=4cm\epsfbox{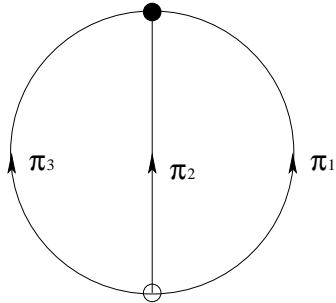}}\end{center}
\caption{\small Two loop vacuum diagram. The filled circle denotes a
twisted vertex and the unfilled circle an untwisted one.}
\label{vacuum}
\end{figure}

Similarly as \cite{Minwalla},
the two loop diagram is evaluated to be
\bea
\label{vacuumtwoloop}
 &&{\rm (bosonic\,\, part)} \otimes \int d^2 \pi_1 d^2 \pi_2 d^2 \pi_3
   \exp\left(-(\pi_1 \wedge \pi_2 +\pi_2\wedge \pi_3+ \pi_1 \wedge \pi_3)
  \right)
  \delta^2(\pi_1+\pi_2+\pi_3)  \nn
 &&= {\rm (bosonic\,\, part)} \otimes {1 \over 4} \det C \delta(0),
\eea
where bosonic part is to be
\be
 {\rm (bosonic\,\, part)} \sim g^2 \bar m^3 \Lambda^2.
\ee
$\Lambda$ is cutoff. But since $\delta(0)=0$, (\ref{vacuumtwoloop}) vanishes.
\begin{figure}[h]
\begin{center}\mbox{\epsfysize=4cm\epsfbox{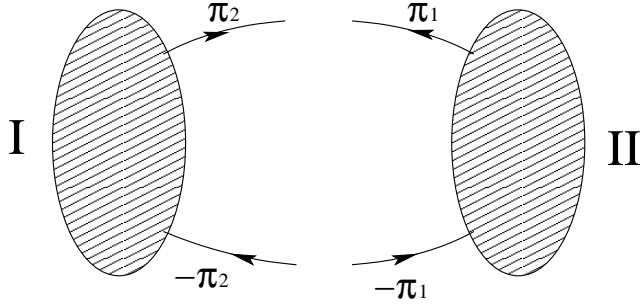}}\end{center}
\caption{\small We have cut general vacuum diagrams into two parts.
}
\label{vacuumgeneral}
\end{figure}

And this feature continues even when we compute arbitrary higher order loops.
The general structure of vacuum diagram is like Fig. \ref{vacuumgeneral}.
We have cut intermediate lines. Since I or II can be a single line, the
cutting is most general. When we join again I and II together to make
a single vacuum diagrams, we need to attach $\delta(\pi_1+\pi_2)$ and
$\delta(-\pi_1-\pi_2)$ at each junction. This gives square of delta function
in the form of $\delta^2(\pi_0+ \cdots)$, where $\pi_0$ is a
loop momenta to be integrated. After integrating out $\pi_0$
this factor gives $\delta(0)$ term which is nothing but zero.
Stated differently, this
is just because there is no single source nor sink of the momentum
flow in the closed vacuum diagram.
 Note that this argument is not true if we attach
additional external lines to the vacuum diagram as tadpole diagrams.
Anyway, we can conclude that the vacuum energy vanishes up to
all higher orders just as $C_{\alpha\beta}=0$ case.
What is surprising here is that
there is no algebraical reason for the vacuum energy to vanish.
Even though
$Q|0\rangle=0$, it does not guarantee the vanishing of
the vacuum energy since $\bar Q$ is a broken generator.

\subsection{One loop diagrams}

Planar diagrams vanish by symmetry just as $C_{\alpha\beta}=0$ case.
Thus we only need to concentrate on nonplanar diagrams.
First, we consider explicitly two point and three point vertex functions
which are directly related to the effective action.
\begin{figure}[h]
\begin{center}\mbox{\epsfysize=4cm\epsfbox{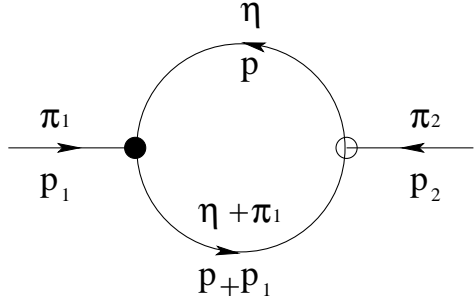}}\end{center}
\caption{\small One loop nonplanar diagram with two external lines. Fermionic
momenta are denoted in Greek and bosonic momenta in Latin.}
\label{onelooptwo}
\end{figure}

Fermionic parts are integrated into
\bea
 \Gamma_{2,F} &=& \int d^2 \eta \exp\left(-{1\over2} \eta \wedge \pi_1\right)
    \exp\left({1\over2} (\eta+\pi_1)\wedge \pi_2\right) \delta(\pi_1+\pi_2) \nn
 &=& {1\over4}
 (\det C) (\pi_1)^2  \delta(\pi_1+\pi_2),
\eea
which does not vanish unless $C_{\alpha\beta}=0$.
Note that the fermionic momentum conservation comes from
$\delta(\pi_1+\pi_2)\equiv(\pi_1+\pi_2)^2$.
Bosonic part integration gives
\bea
 \Gamma_{2,B} &=& {\bar m}^2 \int {d^4 p \over (2\pi)^4}
  {1 \over p^2+m \bar m}
  {1 \over (p+p_1)^2 + m \bar m}  \nn
 &=& {{\bar m}^2 \over (4\pi)^2} \left( \log{\Lambda^2 \over m \bar m}
 + \cdots \right).
\eea
One can see that bosonic loop integration is logarithmically divergent.
The absence of higher order divergence is the consequence of still-remaining
${\cal N}=1/2$ supersymmetry.
Combining both bosonic and fermionic parts together, we get
\be
\label{twopointeffective}
 S_{\rm eff} = \int d^4 x d^2 \pi_1 d^2 \pi_2 {1\over 4} \left({\bar m^2 \over
4 \pi^2} \log {\Lambda^2\over m\bar m} + \cdots \right) \det C
 \Phi(x, \pi_1) (\pi_1)^2 \Phi(x, \pi_2) \delta(\pi_1+\pi_2).
\ee
We note that $(\pi_1)^2$ corresponds to $Q^2$ in $\theta$ space.
Schematically this term can be written as
\be
 S \sim \int d^4 x d^2 \theta \Phi Q^2 \Phi.
\ee
This is still chiral since both $\Phi$ and $ Q^2 \Phi$ are chiral superfields.
We see this is a quantum mechanically induced term which is not
present in the original classical action. To make the theory
renormalizable, we need to add the counter term to cancel this
logarithmic divergence. Stated differently, the
tree level deformed Wess-Zumino action is not enough quantum mechanically,
but we should extend it to accommodate
$\Phi Q^2 \Phi$ term which cannot be deduced from the $*$ deformations 
of ${\cal N}=1$ supersymmetric action.

Also note that, since the induced term $\int d^2 \theta \Phi Q^2 \Phi$ 
gives mass splitting, it is evident that
the mass of the fermion and boson is no longer the same.
This is another novel feature of the deformed Wess-Zumino model. 
Phenomenological application of this feature will be an interesting
subject of future study. 

\begin{figure}[h]
\begin{center}\mbox{\epsfysize=5cm\epsfbox{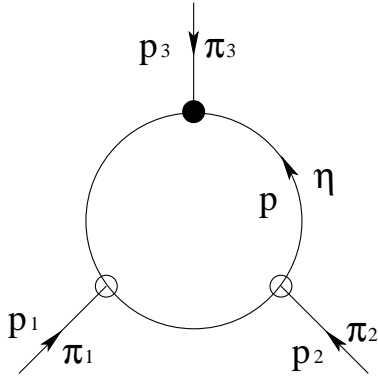}}\end{center}
\caption{\small One loop nonplanar diagrams with three external lines.}
\label{oneloopthree}
\end{figure}

Three point function is to be evaluated similarly. The fermionic part
integration is given by
\be
 \Gamma_F = \left( \exp (-{1 \over 2} \pi_1 \wedge \pi_2 ) +
  \exp (+{1 \over 2} \pi_1 \wedge \pi_2 ) \right) {1 \over 4} \pi_3 \pi_3
 \det C \delta(\pi_1+\pi_2+\pi_3).
\ee
We see that the phase factor can be expressed as $\Phi * \Phi $ and
the the result of loop integration is $\sim \pi \pi$ as
that of two point function.  Bosonic part integration is
\be
 \left({g \over 3}\right)^3 \int {d^4 \over (2 \pi)^4}
 {\bar m \over p^2 + m\bar m}
{\bar m \over (p+ p_1)^2 + m\bar m}
{\bar m \over (p+ p_1+ p_2)^2 + m\bar m},
\ee
 which is finite.  Thus we don't need any further
counter terms to cancel infinities. We can see easily that
four and higher order point functions are also finite since
the loop integration is just the same  as
scalar $\phi^3$ theory.

General structure of fermionic integration of N point function
is summarized as
\bea
 \Gamma_{N,F} &=& \frac{1}{2}
\int d^2 \pi_0 \sum_{i,j=1}^{n} \exp(-{1\over2} \pi_i\wedge\pi_j)
  \delta(\sum_{i=1}^n \pi_i -\pi_0) \otimes
  \left( {1 \over 4} (\pi_0)^2 \det C \right)\nn
  && \,\,\,
   \otimes \sum_{i,j=n+1}^{N-n} \exp(-{1\over2} \pi_i\wedge\pi_j)
  \delta(\sum_{i=1}^n \pi_i +\pi_0).
\eea
We see that novel factorization property here. This is the unique
property of star product, which is also true for bosonic noncommutativity
\cite{Yee}.

If we take $\bar m \rightarrow \infty $ limit
of the action (\ref{original}), bosonic part of the
loop integral is simplified quite a lot.
\be
 \left({g\over 3}\right)^N \int {d^4 p \over (2\pi)^4} \left({1\over m}
 \right)^N \sim \left({g \over m} \right) ^N \Lambda^4
\ee
But, this term is badly divergent. Moreover since all N point functions have
this divergence, this theory is nonrenormalizable.
But in this case, the effective action is factorized into a very simple form,
because loop integrations are factored out.
The effective action is resummed to be
\be
 S_{\rm eff} = \frac{1}{2}
\int d^4 x d^2 \pi_0 \exp\left({g \over 3m} \Phi(x,\pi_0)\right)_*
  \left( {1\over 4} \pi_0 \pi_0 \det C  \Lambda^4 \right)
   \exp\left({g \over 3m} \Phi(x,-\pi_0)\right)_*,
\ee
where
\beq
\exp\left(\Phi(x,\pi)\right)_* \equiv
\sum_{l=1}^{\infty} \frac{1}{l !}
\int d^2 \theta e^{\theta \pi}
\left( \Phi(x,\theta) \right)^l_*
\eeq
This is reminiscent to the expression of the 1-loop effective action of
bosonic noncommutative field theories, where effective actions can be
resummed as interactions between two open Wilson lines\cite{Yee}.
Even though this structure looks interesting with relation to
the open-closed string duality, further study is needed to understand
the full meaning in terms of string theory.

\subsection{Component fields}

If $\bar g=0$ limit is taken, the potential (\ref{potentialwithoutF}) becomes
\be
  V= m\bar m A \bar A + g\bar m A^2\bar A
  + {1\over 3} g \bar m^3 \det C \bar A^3.
\ee

\begin{figure}[h]
\begin{center}\mbox{\epsfysize=3.5cm\epsfbox{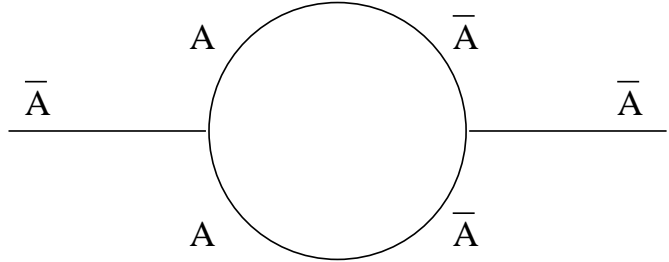}}\end{center}
\caption{\small $\bar A \bar A$ term which is induced quantum mechanically.}
\label{component}
\end{figure}

 We can see very easily that this potential induces a quantum mechanical
term
\be
 S_{\rm eff}  \sim \int d^4 x \bar A(x) \bar A(x),
\ee
which can be expressed in terms of superfield
\bea
 S_{\rm eff}
   &\sim& \int d^4 x d^2 \theta \Phi(x,\theta) Q^2 \Phi(x, \theta).
\eea

Thus again we see that $\int \Phi Q^2 \Phi$ is needed for quantum completeness.

\subsection{Tadpole contributions}

We have seen that vacuum diagrams vanish in general even though it is
nonplanar. This is caused by momentum conservation  $\delta^2(\cdots)$
factor attached. But for tadpole diagrams, there is  no
more $\delta^2(\cdots)$ factor since tadpoles have outer source for
momentum flow.
 Thus nonplanar tadpole diagrams do not vanish in general.
We will present explicit examples here.

\begin{figure}[h]
\begin{center}\mbox{\epsfysize=4cm\epsfbox{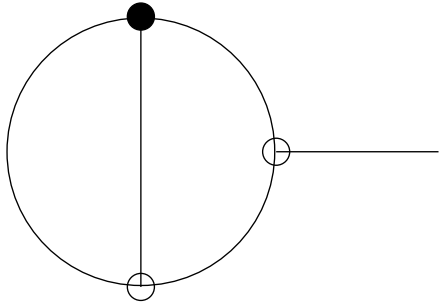}}\end{center}
\caption{\small An example of nonplanar tadpole diagram.}
\label{tadpole}
\end{figure}

Bosonic loop integration gives $g^3 \bar m^4 \log{\Lambda^2 \over m \bar m}$.
With fermionic integration included, the effective action is
\be
 S_{\rm eff} \sim \int d^4 x d^2 \pi_0
 \left(g^3 \bar m^4 \log{\Lambda^2 \over m \bar m}
 \right) \det C \delta (\pi_0) \Phi(x,\pi_0).
\ee
Generally speaking, we need redefinition of fields to cancel
liner terms. Usually  such kinds of field redefinition
give nonzero cosmological constant. But will it be the case also here?

This tadpole term amounts to adding $S F$ in componentwise to the
classical potential (\ref{potential}). Here
 $S \sim g^3 \bar m^4 \log{\Lambda^2 \over m \bar m}
 \det C $. Then the general form of the modified potential becomes
\be
 V_{\rm modified} = - F \bar F - F ( m A + g A^2 + S)
                    - H(F) - {g \over 3} \det C F^3 -
                    \bar m \bar A \bar F.
\ee
Here we added a term $ H(F)$  coming from one loop computation since
we are calculating two loop diagrams.
Note that $H(F)$ is a function of $F$ only and there is no
$\bar F$ correction. It is easy to see that, 
to eliminate the linear term, we need
to redefine $A \rightarrow A + T$ where
$T$ satisfies the equation $g T^2 + m T + S=0$.  
This equations always has a solution.
Thus the tadpole term does not induce any cosmological constant.
We should note that this is true because only $A$ needs to be redefined
but $\bar A$ remain intact. This is possible
because we are treating Euclidean theories.

\section{Discussions}

Considering quantum fluctuations of field theories in noncommutative superspace,
we get additional terms as $\int d^2 \theta \Phi Q^2 \Phi$ which is not
present in the original $*$-deformed Wess-Zumino model.
Thus we can say that the star deformation
of the noncommutative superspace is not quantum mechanically complete,
but we need to extend the model to accommodate such operators.
We have shown the possibility of this extension
since the set of ${\cal N}=1/2$ supersymmetric action in general
is larger than the set of $*$-deformation
of ${\cal N}=1$ supersymmetric action.
{}From the string theory point of view, it is quite interesting
to ask whether such kind of extension is also natural when we
consider gauge theories  in the noncommutative superspace
derived from string theory. Especially it will be interesting
to trace the the possible stringy origin of the terms
induced by quantum fluctuation.

Dijkgraaf-Vafa theory \cite{DiVa1, DiVa2}
can be considered in this setting,
since even though there is no ``holomorphicity'', there is the notion of
``anti-holomorphicity''.
Actually, considering the superpotential
for the case of $\bar{g}=0$,
we see that there is no $\bar{m}$ dependence for the planar diagrams.
The bosonic and fermionic integrations cancel each other. But as 
we have shown explicitly in the section 3, the
contributions from nonplanar graphs are not zero. Since there is
no cancellation between bosonic and fermionic integrations,
the effective superpotential depends on $\bar{m}$. Thus to fully understand
the quantum structure including nonplanar diagrams, 
we need further study in this direction.

\vskip6mm

\noindent
{\bf Acknowledgment}

We thank Jan de Boer for discussions.
This work is supported in part by the Stichting FOM. \\

\noindent
{\bf Note added}:

As this article was being completed,
we received the preprint
\cite{Rey} which partly overlaps with the present work.

\appendix
\setcounter{equation}{0}

\newpage



\begin{thebibliography}{99}


\bibitem{SeWi}
N.~Seiberg and E.~Witten,
``String theory and noncommutative geometry,''
JHEP {\bf 9909} (1999) 032
[arXiv:hep-th/9908142].



\bibitem{FeLl}
S.~Ferrara and M.~A.~Lledo,
``Some aspects of deformations of supersymmetric field theories,''
JHEP {\bf 0005} (2000) 008
[arXiv:hep-th/0002084].

\bibitem{Te}
S.~Terashima,
``A note on superfields and noncommutative geometry,''
Phys.\ Lett.\ B {\bf 482} (2000) 276
[arXiv:hep-th/0002119].


\bibitem{Ca}
R.~Casalbuoni,
``Relativity And Supersymmetries,''
Phys.\ Lett.\ B {\bf 62} (1976) 49.


\bibitem{Casalbuoni:1975bj}
R.~Casalbuoni,
``On The Quantization Of Systems With Anticommutating Variables,''
Nuovo Cim.\ A {\bf 33} (1976) 115.

\bibitem{Casalbuoni:1976tz}
R.~Casalbuoni,
``The Classical Mechanics For Bose-Fermi Systems,''
Nuovo Cim.\ A {\bf 33} (1976) 389.

\bibitem{Schwarz:pf}
J.~H.~Schwarz and P.~Van Nieuwenhuizen,
``Speculations Concerning A Fermionic Substructure Of Space-Time,''
Lett.\ Nuovo Cim.\  {\bf 34} (1982) 21.


\bibitem{Klemm:2001yu}
D.~Klemm, S.~Penati and L.~Tamassia,
``Non(anti)commutative superspace,''
Class.\ Quant.\ Grav.\  {\bf 20} (2003) 2905
[arXiv:hep-th/0104190].


\bibitem{Abbaspur:2002xj}
R.~Abbaspur,
``Generalized noncommutative supersymmetry from a new gauge symmetry,''
arXiv:hep-th/0206170.




\bibitem{deBoer}
J.~de Boer, P.~A.~Grassi and P.~van Nieuwenhuizen,
``Non-commutative superspace from string theory,''
arXiv:hep-th/0302078.

\bibitem{Ooguri1}
H.~Ooguri and C.~Vafa,
``Gravity induced C-deformation,''
arXiv:hep-th/0303063.

\bibitem{Ooguri2}
H.~Ooguri and C.~Vafa,
``The C-deformation of gluino and non-planar diagrams,''
arXiv:hep-th/0302109.

\bibitem{Kawai}
H.~Kawai, T.~Kuroki and T.~Morita,
``Dijkgraaf-Vafa theory as large-N reduction,''
arXiv:hep-th/0303210.


\bibitem{Chepelev:2003ga}
I.~Chepelev and C.~Ciocarlie,
``A path integral approach to noncommutative superspace,''
arXiv:hep-th/0304118.


\bibitem{David}
J.~R.~David, E.~Gava and K.~S.~Narain,
``Konishi anomaly approach to gravitational F-terms,''
arXiv:hep-th/0304227.







\bibitem{Seiberg}
N.~Seiberg,
``Noncommutative superspace, N = 1/2 supersymmetry, field theory and  string theory,''
JHEP {\bf 0306} (2003) 010
[arXiv:hep-th/0305248].


\bibitem{BeSe}
N.~Berkovits and N.~Seiberg,
``Superstrings in Graviphoton Background and N=1/2+3/2 Supersymmetry,''
arXiv:hep-th/0306226.


\bibitem{AsSuTe}
T.~Asakawa, S.~Sugimoto and S.~Terashima,
``Exact description of D-branes via tachyon condensation,''
JHEP {\bf 0302} (2003) 011
[arXiv:hep-th/0212188].




\bibitem{SeIn}
K.~A.~Intriligator and N.~Seiberg,
``Lectures on supersymmetric gauge theories and electric-magnetic  duality,''
Nucl.\ Phys.\ Proc.\ Suppl.\  {\bf 45BC} (1996) 1
[arXiv:hep-th/9509066].

\bibitem{DiVa1}
R.~Dijkgraaf and C.~Vafa,
``Matrix models, topological strings, and supersymmetric gauge theories,''
Nucl.\ Phys.\ B {\bf 644} (2002) 3
[arXiv:hep-th/0206255].

\bibitem{DiVa2}
R.~Dijkgraaf and C.~Vafa,
``A perturbative window into non-perturbative physics,''
arXiv:hep-th/0208048.


\bibitem{Zanon}
R.~Dijkgraaf, M.~T.~Grisaru, C.~S.~Lam, C.~Vafa and D.~Zanon,
``Perturbative computation of glueball superpotentials,''
arXiv:hep-th/0211017.

\bibitem{Minwalla}
S.~Minwalla, M.~Van Raamsdonk and N.~Seiberg,
``Noncommutative perturbative dynamics,''
JHEP {\bf 0002} (2000) 020
[arXiv:hep-th/9912072].

\bibitem{Yee}
Y.~Kiem, S.~J.~Rey, H.~T.~Sato and J.~T.~Yee,
``Open Wilson lines and generalized star product in nocommutative scalar  field theories,''
Phys.\ Rev.\ D {\bf 65} (2002) 026002
[arXiv:hep-th/0106121].




\bibitem{Rey}
R. ~Britto, B.~Feng and S.-J.~Rey,
``Deformed Superspace, N=1/2 Supersymmetry and (Non)Renormalization Theorems,''
arXiv:hep-th/0306215.


\end{thebibliography}
\end{document}